\documentclass{ws-ijmpa}

\begin{document}


\catchline{}{}{}

\title{ THE SCALAR ETHER-THEORY OF GRAVITATION\\ AND
 ITS FIRST TEST IN CELESTIAL MECHANICS }

\author{\footnotesize MAYEUL ARMINJON}

\address{ Laboratoire ``Sols, Solides, Structures" [U.M.R. 5521 of
the CNRS]\\ B.P. 53, 38041 Grenoble cedex 9, France. }

\maketitle

\pub{Received (Day Month Year)}{Revised (Day Month Year)}

\begin{abstract}
The motivations for investigating a theory of gravitation based on
a concept of ``ether" are discussed-- a crucial point is the
existence of an alternative interpretation of special relativity,
named the Lorentz-Poincar\'e ether theory. The basic equations of
one such theory of gravity, based on just one scalar field, are
presented. To check this theory in celestial mechanics, an
``asymptotic" scheme of post-Newtonian (PN) approximation is
summarized and its difference with the standard PN scheme is
emphasized. The derivation  of PN equations of motion for the mass
centers, based on the asymptotic scheme, is outlined. They are
implemented for the major bodies of the solar system  and the
prediction for Mercury is compared with an ephemeris based on
general relativity. 
\end{abstract}

\section{ Introduction: Why an Ether Theory of Gravitation?}

(i) Our first motivation was to extend the {\it Lorentz-Poincar\'e
ether theory} so that gravitation be included. The
Lorentz-Poincar\'e ether theory may be described as the theory
according to which a) the ether is an inertial frame E such that
Maxwell's equations are valid in E, and b) any material object
that moves with respect to E undergoes a Lorentz contraction. As
shown, in particular, by Prokhovnik,\cite{Prokhovnik} this theory
is physically equivalent to standard {\it special} relativity
(SR). This makes its ether undetectable, hence ``superfluous"
(Einstein 1905). But since SR does not involve gravitation, we may
ask whether a preferred frame could exist but remain hidden in the
absence of gravitation, and become detectable in its presence.\\

(ii) An ``ether" could help to make quantum theory and gravitation
theory compatible. Quantum theory was originally built in a flat
space-time, moreover it uses a preferred time. In flat space-time,
this is the inertial time, which depends on the inertial frame in
a way that remains compatible with Lorentz invariance. Yet in the
curved space-time of gravitation, there seems to be no way to
prefer some time coordinate, except if we {\it a priori} admit a
``preferred space-time foliation", i.e. an ether. This may be
illustrated already for the case of the Klein-Gordon equation, for
which many possible extensions to curved space-time {\it a priori}
exist, but one is preferred if we have an ether.\cite{Arm98a}
Further, quantum theory shows that ``vacuum" has physical effects,
e.g. the Casimir effect, now experimentally confirmed.\\

(iii) Investigating a strongly alternative theory also opens a new
way to solve some problems common to general relativity (GR) and
to most extensions of it: a) Singularities: The investigated
``scalar ether-theory" does avoid singularities, in gravitational
collapse\cite{Arm97} and in cosmology as well.\cite{Arm01a} b)
Gauge condition: The solutions to the underdeterminacy of the
Einstein equations as a system of partial differential equations
are either to say that the Lorentz manifold is determined modulo
diffeomorphisms, or to add a gauge condition in a fixed space-time
manifold (the latter way is used in applications). What is the
precise link between these two solutions? In the scalar
ether-theory, there is no need for any gauge, yet space-time is
fixed. c) Galactical dark matter: Identified candidates seem
poorly found. In the present ether-theory, the preferred-frame
effects are probably more important at the galactical scale and
beyond, due to the large time scales involved.

\section{Basic Equations of the Investigated Theory}

N.B.: Most equations are preferred-frame ones with space
covariance only.\\

(i) Gravitation is seen as  Archimedes' thrust in an imagined
perfect fluid (``ether") with pressure $p_e$ and density $\rho_e =
\rho_e(p_e)$. This leads to define the {\it gravity acceleration
vector} as follows:\cite{Arm97}
\begin{equation} \label{gvector}
  \mathbf{g}=-\frac{\mathrm{grad}~p_e}{\rho_e}.
\end{equation}
Note that, due to Eq.~(\ref{gvector}),  $p_e$ and $\rho_e$
decrease towards the attraction, thus $\rho_e(\mathbf{x},t) <
\rho_{e}^{\infty}(t) \equiv \mathrm{Sup}_{\mathbf{x}\in
\mathrm{M}} \rho_e(\mathbf{x},t)$ in a gravitational field, where
M is the ``space" manifold, i.e. the set of the positions
$\mathbf{x}$ in the preferred frame E.\\

(ii) {\it Assumed metric effects of a gravitational
field.}\cite{Arm97} We assume that the space-time
$\mathbf{R}\times\mathrm{M}$ is equipped with a flat metric
$\mathbf\gamma^0$ for which the preferred frame E is an inertial
(Galilean) frame. The inertial time $t$ in E is called the
``absolute time", and the Euclidean space metric associated with
$\mathbf\gamma^0$ in the frame E is denoted by
$\mathbf\mathsf{g}^0$. Yet we also assume that, in a gravitational
field, i.e. $\rho_e(\mathbf{x},t) < \rho_{e}^{\infty}(t)$, there
are metric effects, similar to those due to uniform motion (see
Ref. 1): the meters are contracted and the clocks are slowed down,
in the ratio $\beta\equiv
\rho_{e}(\mathbf{x},t)/\rho_{e}^{\infty}(t)$ and (for the meters)
in the direction $\mathbf{g}$ only. This means a dilation
(contraction) of the length (time) intervals, when they are indeed
measured with physical instruments, as compared with those that
would be evaluated in terms of the flat space metric
$\mathbf\mathsf{g}^0$ and the absolute time $t$. Hence, the
``physical" space metric $\mathbf\mathsf{g}$ in the frame E is a
Riemannian one, and the measured time is a ``local" one, denoted
by $t_{\mathbf{x}}$ (at point $\mathbf{x}\in \mathrm{M}$). Thus,
the ``physical" space-time metric $\mathbf\gamma$ is a curved
Lorentzian metric. Moreover, SR leads to assuming the relation
$p_e=c^2 \rho_e$. \\

 (iii) {\it Gravitational field equation.} The following equation is stated
 for the scalar gravitational field $p_e$:
\begin{equation} \label{field}
\Delta_{\mathbf\mathsf{g}} p_e - \frac{1}{c^2} \frac{\partial^2
p_e }{\partial t_{\mathbf{x}}^2}= 4 \pi G \sigma \rho_e,
\end{equation}
where $\sigma$ is the $T^{00}$ component of the energy-momentum
tensor of matter and nongravitational fields {\bf T}, when the
time coordinate is $x^0=ct$ with $t$ the absolute time, and in any
spatial coordinates that are adapted to the preferred reference
frame E.\cite{Arm96} The derivative with respect to the local time
is defined by
\begin{equation}\label{localtime}
 \frac{ \partial}{\partial t_\mathbf{x}}=
 \frac{1}{\beta(t,\mathbf{x})}\! \frac{\partial}{\partial t}.
\end{equation}
In Eq.~(\ref{field}), the Laplace(-Beltrami) operator is defined
with the curved space metric $\mathbf\mathsf{g}$ (relative to the
frame E). The same is true for the $\mathrm{grad}$ operator in
Eq.~(\ref{gvector}).\\

(iv) {\it Dynamics is governed by Newton's second law:} force =
time-derivative of momentum.\cite{Arm96} The force over the test
particle is the gravitational force $m(v)\mathbf{g}$, plus the
nongravitational (e.g. electromagnetic) force, where $m(v)$ is the
relativistic inertial mass, involving the Lorentz factor. The
momentum is $m(v)\mathbf{v}$; $\mathbf{v}$ and $v = |\mathbf{v}|$
are evaluated with the physical metric. The time-derivative of
momentum is uniquely defined from compelling requirements
(including Leibniz' rule for a scalar product).\cite{Arm96} In the
{\it static} case, that extension of Newton's 2nd law {\it implies
Einstein's geodesic motion.} For a dust, we may apply this
extension pointwise in the continuum, and it implies a new
equation for continuum dynamics:\cite{Arm98b}

\begin{equation} \label{dynamics}
T_{\mu ;\nu}^\nu = b_\mu, \; b_0 \equiv \frac{1}{2}g_{jk,0}T^{jk},
\: b_i \equiv - \frac{1}{2}g_{ik,0}T^{0k}.
\end{equation}
The universality of gravity is expressed in the fact that
Eq.~(\ref{dynamics}) is assumed to hold true for any material
medium (thus also for a nongravitational field).

\section{Asymptotic Post-Newtonian Approximation}

An ``asymptotic" post-Newtonian approximation (PNA) was
developed\cite{Arm00a} (cf. Futamase \& Schutz\cite{FutaSchutz}
and Rendall\cite{Rendall92} in GR; in GR, the local field
equations of the asymptotic method have not been used to get
equations of motion for the mass centers of extended bodies. This
has beeen done in the present theory\cite{Arm00b,Arm02b}):\\

(i) The gravitational field {\it and} the matter fields are
expanded (in the standard PNA,\cite{Fock64,Chandra65} only the
gravitational field is expanded).\\

(ii) For definiteness, each body is assumed to be made of a
barotropic perfect fluid (one fluid per body). Other constitutive
laws may also be considered, of course.\\

(iii) A family ($\mathrm{S}_\lambda$) of gravitating systems is
deduced from the given system S. To do this, we use the fact that
an exact similarity transformation exists in Newtonian
gravity.\cite{Arm00a,FutaSchutz} This transformation is applied to
the initial data for S. For this, $V \equiv c^2(1-f)/2$ is
substituted for the Newtonian potential, where $f \equiv
\gamma_{00}$ ($1-f \ll 1$ for a weak field). The initial data for
S (the system of interest, e.g. the solar system) is
general,\cite{Arm00a} in contrast with Ref.~8 in which the initial
space metric was very special. (In Ref.~9, the family was {\it a
priori} assumed.)
\\

Adopting units $[\mathrm{T}]_\lambda = [\mathrm{T}]/\lambda^{1/2}$
and $[\mathrm{M}]_\lambda = \lambda[\mathrm{M}]$ for system
$\mathrm{S}_\lambda$, all fields are ord($\lambda^0$), and the
small parameter $\lambda$ is proportional to $1/c^2$ (in fact
$\lambda = (c_0/c)^2$, where $c_0$ is the velocity of light in the
starting units [T] and [M]). Therefore, the derivation of
asymptotic expansions is straightforward.\cite{Arm00a} (In the
standard PN scheme, $1/c^2$ is {\it formally} considered as a
small parameter.) That $1/c^2$, not $1/c$, turns out to be the
effective small parameter, is due to the fact that it is only
$1/c^2$ that enters in the equations. The theory admits consistent
expansions in powers of $\lambda$ (or $1/c^2$). The first
(zero-order) term is Newtonian gravity, hence the theory admits a
correct Newtonian limit. The first-order approximation in
$\lambda$ or $1/c^2$ is the first PN approximation. Using these
expansions is justified insofar as the system of interest
corresponds to a small value $\lambda_0$ of $\lambda$, which is
the case, e.g., for the solar system.

\section{PN Equations of Motion for the Mass Centers}

The mass centers (MC) are defined\cite{Arm00b} as local
barycenters of the rest-mass density $\rho_{\mathrm{exact}}$
(instead of, e.g., the active energy density $\sigma$,
Eq.~(\ref{field})), because (i) $\rho_{\mathrm{exact}}=0$ outside
the bodies (which is wrong if one takes instead a density that
involves gravitational energy, since the latter is distributed in
the whole space) and (ii) $\rho_{\mathrm{exact}}$, or rather its
PN approximation, obeys the usual continuity equation, i.e.
without adding gravitational energy and its flux. To get the MC's
equation of motion, one just integrates the local equations of
motion inside the different bodies.\cite{Arm00b,Arm02b} Due to the
use of the asymptotic method, the local equations of orders 0 and
1 in $\lambda$ or $1/c^2$ are separated. E.g.:

\begin{equation}\label{expandcontinuity}
 \partial_{t}\rho_{0} + \partial_{j}(\rho_{0}u_{0}^j) = 0,~~
 \partial_{t}\rho_{1} + \partial_{j}(\rho_{1}u_{0}^{j}+\rho_{0}u_{1}^{j}) =
0
\end{equation}
for the continuity equation, derived from the time component at
the first PNA. The equation of order 1 is linear with respect to
the fields of order 1. Therefore, separate equations are also
obtained for the MC's, and the equation for PN corrections (order
1) is linear with respect to order-1 quantities. To get tractable
equations, every field is decomposed into a self-field and an
external field, and account is taken of the ``good separation"
between different bodies, which means that
\begin{equation}\label{def_eta}
\eta\equiv\mathrm{Sup}_{a\neq
b}(r_{a}/|\mathbf{x}_{a}-\mathbf{x}_{b}|)\ll 1
\end{equation}
($r_a$ is the radius of body ($a$)). In the solar system, terms up
to and including $\eta^{3}$ must be retained.\cite{Arm02b} A rigid
motion, possibly including self-rotation, is assumed for each
body. Finally, the rest-mass density of the order 0, $\rho_{0}$,
is assumed spherical for each body, at the stage of calculating
the PN {\it corrections}. One thus gets explicit equations of
motion for the mass centers.\cite{Arm02b} They show that the
self-rotation of the bodies {\it and their internal structure}
influence the motion from the first PNA. {\it This follows
naturally from using the asymptotic method and should hold true
for GR.}

\section{Implementation. Comparison with a General-Relativistic Ephemeris}

In order to use the equations of motion for the mass
centers\cite{Arm02b} so as to check the theory, we have to know
the values of the parameters that enter these equations. These are
the 0-order masses $M_a$ of the bodies (here the major bodies of
the solar system), the initial conditions of their motion, and the
constant velocity $\mathbf{V}$ of the global zero-order mass
center of the solar system, with respect to the preferred frame E
(and also the constant $G$).\cite{Arm00b} (Of course there is no
parameter like $\mathbf{V}$ in conventional theories.) These
unknown parameters depend on the theory. They must be determined
by optimizing the agreement between predictions and
observations.\cite{Arm00b} Our computer code loops on the
numerical solution of the translational equations of motion in
order to optimize the parameters.\cite{Arm02c} This code has been
tested by investigating in which measure one may reproduce (over
one century) the predictions of the DE403
ephemeris,\cite{Standish95} by using purely Newtonian equations of
motion.\cite{Arm02c} It has also been applied to adjust over 60
centuries a less simplified model, in which the PN corrections in
the Schwarzschild field of the Sun are also
considered.\cite{Arm02a}\\
\begin{figure}
  \centering
  \includegraphics{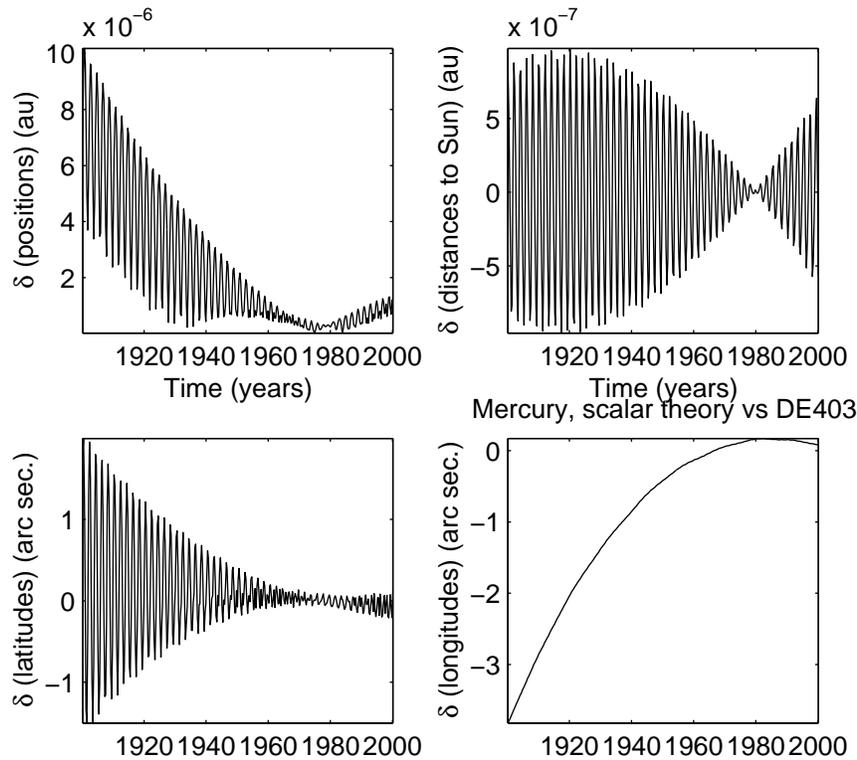}
  \caption{Difference between the scalar theory and the DE403 ephemeris of the JPL.}
  \label{1}
\end{figure}

In the version of the code that incorporates the equations of
motion\cite{Arm02b} derived from the present theory, a Lorentz
transform allows to pass from the preferred reference frame to the
frame bound with the zero-order global barycenter, and vice-versa.
This transform, as well as the inverse transform, is determined by
the adjustable vector $\mathbf{V}$. Thus, the adjustment process
of the translational equations on observational data provides us
eventually with the value of $\mathbf{V}$ that minimizes the
residual with the set of observations. Note that the
``observational data" are currently taken from an {\it ephemeris
based on GR,}\cite{Standish95} specifically we take a set of
heliocentric positions of the eight major planets, between 1956
and 2000. With these input data, themselves a fitting of
observations by GR equations, the magnitude of the optimal vector
$\mathbf{V}$ is $|\mathbf{V}|\simeq$ 3 km/s, which is significant.
The difference between DE403 and our thus-adjusted equations of
motion is small (cf. the residual advance in Mercury's longitude
of perihelion, with respect to Newton's theory: 43"), but
significant (Fig. 1. The self-rotation of all nine bodies is
neglected). Our current project is to adjust the theory on a set
of true {\it astronomical observations.}

\end{document}